\documentclass[11pt]{article}
\usepackage[margin=1.2 in]{geometry}  
\usepackage{graphicx}
\usepackage{amssymb,amsmath}
\usepackage[numbers,sort&compress]{natbib}
\usepackage{array}
\usepackage{multirow}
\usepackage[shortlabels]{enumitem}
\usepackage[dvipsnames]{xcolor}
\usepackage{hyperref}

\usepackage{algorithm}
\usepackage[noend]{algpseudocode}

\usepackage[charter]{mathdesign}
\usepackage[mathcal]{eucal}

\usepackage{tikz}
\usetikzlibrary{arrows.meta} 
\tikzset{
    vertex/.style={circle, draw, fill=black!50, inner sep=0pt, minimum width=5pt}
}
\usetikzlibrary{calc}
\usepackage{pgfplots}
\pgfplotsset{compat=1.18}

\newcommand{\dd}[2]{\frac{d #1}{d #2}}

\newcommand{\grad}{\nabla}
\newcommand{\half}{\frac{1}{2}}

\newcommand{\R}{\mathbb{R}}

\newcommand{\e}[1]{^{(#1)}}
\newcommand{\new}{_{{\rm new}}}
\newcommand{\sym}{_{{\rm sym}}}
\newcommand{\chol}{_{{\rm chol}}}
\newcommand{\lu}{_{{\rm LU}}}
\newcommand{\iter}{_{{\rm iter}}}
\newcommand{\est}{^{{\rm est}}}

\title{Monte Carlo on manifolds in high dimensions}
\author{Kerun Xu$^1$ and Miranda Holmes-Cerfon$^2$}
\date{\small %
    $^1$Courant Institute of Mathematical Sciences, New York University\\%
    $^2$ Department of Mathematics, University of British Columbia and \\Courant Institute of Mathematical Sciences, New York University. \\1984 Mathematics Rd, Vancouver, BC, V6T 1Z2, \href{}{holmescerfon@math.ubc.ca}  
    \\[2ex]%
    \today
}

\begin{document}

\maketitle










\begin{abstract}
    We introduce an efficient numerical implementation of a Markov Chain Monte Carlo method to sample a probability distribution on a manifold (introduced theoretically in Zappa, Holmes-Cerfon, Goodman (2018) \cite{Zappa:2018jy}), where the manifold is defined by the level set of constraint functions, and the probability distribution may involve the pseudodeterminant of the Jacobian of the constraints, as arises in physical sampling problems. The algorithm is easy to implement and scales well to problems with thousands of dimensions and with complex sets of constraints provided their Jacobian retains sparsity. The algorithm uses direct linear algebra and requires a single matrix factorization per proposal point, which enhances its efficiency over previously proposed methods but becomes the computational bottleneck of the algorithm in high dimensions. We test the algorithm on several examples inspired by soft-matter physics and materials science to study its complexity and properties. 
\end{abstract}

\section{Introduction}

Equality constraints occur widely in physical and statistical sampling problems. In a physical system, such as a collection of interacting particles, constraints may be used to remove stiff degrees of freedom \cite{Ryckaert:1977wr,vanGunsteren:1980gp,Hartmann2008,Lelievre:2012id}. For example, a stiff spring between particles may be replaced by a fixed distance between them, which allows taking a longer step on the submanifold that preserves this distance constraint. 
In statistics, constraints or approximate constraints may arise as a condition on the set of parameters in a parametric or Bayesian model \cite{klugkist2007bayes,Diaconis:2013ka,mulder2014,griffiths2020constrained,Beskos2022}, and they may accelerate training of neural networks \cite{leimkuhler2020constraint}.  
In algebraic geometry, constraints may be used to study properties of algebraic varieties, such as surface or boundary integrals \cite{Magron2020}, metric properties \cite{Nassar2013}, or topological properties \cite{Marigliano2020}. 
In all such systems, equality constraints reduce the intrinsic dimension of the space being sampled, 
so the probability measure of interest lives on a submanifold of the ambient space. 

Sampling a measure supported on a submanifold requires special consideration, and a number of ideas and algorithms have been developed to simulate or sample from systems subject to holonomic constraints. These go back to the Shake \cite{Ryckaert:1977wr} and Rattle \cite{Andersen:1983wg} algorithms from molecular dynamics, which are still widely used to this day, and have been extended to a variety of other systems such as Langevin samplers \cite{VandenEijnden:2006gp,Ciccotti:2007fv,Lelievre:2012id,Leimkuhler:2016ia,zhang2020,zhang2021}, Hamiltonian samplers \cite{Brubaker:2012wy,Lelievre:2019be}, geodesic samplers \cite{Byrne:2013ed}, and Monte Carlo samplers \cite{Zappa:2018jy}. 
Yet constraints don't come for free: since constraints are typically nonlinear functions of a system's variables, imposing these constraints requires solving a nonlinear system of equations for the Lagrange multipliers at each step of the algorithm, except for special manifolds for which an explicit parameterization is available \cite{Diaconis:2013ka,Roberts2022stereographic}. Therefore, applying these methods to high-dimensional systems requires carefully considering the numerical aspects of the algorithms. Such consideration has been applied to algorithms in molecular dynamics \cite{Barth:1995vi,Krautler2001,Yoneya:2001ds,Weinbach:2005jt,Gonnet2007,Eastman:2010ts}, which may consider systems with $O(1000+)$ degrees of freedom, but has been less applied to Monte Carlo algorithms, where a slightly different set of issues arise.

Our goal in this paper is to describe an efficient numerical implementation of a Markov Chain Monte Carlo (MCMC) sampling method with constraints, which allows sampling problems with $O(1000)$ dimensions while remaining easy to implement using standard linear algebra packages. 
We directly build on \cite{Zappa:2018jy}, which introduced an MCMC algorithm on manifolds but for which the implementation described in that reference could not scale to high dimensions. 
We focus on physically-motivated problems inspired by soft-matter physics and materials science \cite{Manoharan:2015ko,HolmesCerfon:2016wa,Bertoldi2017} where the constraints are approximations for stiff forces, because these have an additional challenge associated with the presence of an ``effective'' measure on the constraint surface, which can be hard to calculate in high dimensions (and appears to be often neglected in molecular dynamics simulations). However, our algorithm can be applied to systems without this effective measure simply by not including it in the Metropolis ratio. Apart from this effective measure, we mainly concentrate on problems with no additional force field, since we wish to focus on the numerical issues associated with handling constraints, and not on problems such as evaluating forces that are common to both constrained and unconstrained sampling problems.
The implementation we describe requires a single matrix factorization per proposal point, which we demonstrate is the main computational bottleneck of the algorithm for complex constraints. 
The algorithm is feasible for problem sizes for which this factorization is possible. 

A key idea that allows for just a single matrix factorization is to solve the nonlinear constraint equations using a quasi-Newton method, by replacing the exact Jacobian of these equations with a symmetric approximation that doesn't change with the iterations of a single Newton-like solve. Such an idea was first introduced in \cite{Barth:1995vi} for molecular dynamics, and subsequently explored in other molecular dynamics algorithms, such as  Ref. \cite{Weinbach:2005jt}, which used this idea to develop a parallel implementation of SHAKE, Ref. \cite{Gonnet2007}, which used this idea along with a preconditioner to speed up SHAKE, and Ref. \cite{Eastman:2010ts}, which proposed explicitly inverting an approximate Jacobian and using this same inverse throughout the simulation. Notably, while this idea in the context of molecular dynamics leads to a typically modest decrease in total computational time, generally 2-10\% and at most 50\% in the examples considered in \cite{Barth:1995vi}, in our examples it leads to a huge decrease in computational time, a minimum of 50\% but up to a 98\% decrease. We believe this is because the examples considered in molecular dynamics have relatively simple constraints and more complex force fields which are expensive to evaluate, whereas we consider examples with complex constraints and simple force fields so our primary computation is in handling the constraints.   

Here is an overview of the paper. Section \ref{sec:setup} introduces the setup, by explaining the precise problem formulation, including how an effective measure arises from delta-functions in a physical setting where constraints are approximations for stiff forces. It also gives an overview of the sampling algorithm. Section \ref{sec:alg} describes the numerical implementation of the algorithm. Section \ref{sec:results} applies the algorithm to four examples and analyzes its properties, such as computational time and complexity, the nature of the rejection statistics, the proportion of time occupied by matrix factorizations. It also studies the effective diffusivity as a function of the average acceptance ratio, finding that the optimal acceptance ratio depends on dimension for examples with flatter probability distributions. Section \ref{sec:conclusion} summarizes the results and puts them in a broader context. Code which implements the examples in this paper can be found at \cite{gitcode}.

\section{Setup and overview of the Monte Carlo algorithm}\label{sec:setup}

We consider a system described by $n$ variables, $x\in \R^n$, which is subject to $m$ constraints  of the form
$q_i(x) = 0$  $(i=1,\ldots,m)$,
where  $q_i:\R^n\to \R$ are continuously differentiable functions, and $m<n$. 
The system is therefore constrained to a set of the form
\begin{equation}
    M = \{x\in \R^n : q_i(x) = 0, i=1,\ldots,m\}.
\end{equation}
We write $q(x) = (q_1(x),\ldots,q_m(x))^T$ to denote the vector-valued constraint function. 

The geometric nature of $M$ is in part determined by the Jacobian of the constraints. Let 
\begin{equation}
    Q_x = \begin{pmatrix}\grad q_1(x) &\cdots& \grad q_m(x)\end{pmatrix}
\end{equation}
be the $n\times m$ matrix whose columns are the gradients of the constraints (taking gradients to be column vectors). This is the transpose of the Jacobian $\grad q(x) = \frac{\partial q}{\partial x}\Big|_{x}$; we refer to it simply as the Jacobian. We make the following assumption: 
\begin{equation}\label{assumption}
\text{\emph{$Q_x$ has full rank everywhere on $M$}.} 
\end{equation}
This assumption implies, by the inverse function theorem, that $M$ is a differentiable manifold, and its dimension is $d=n-m$. 
The tangent space to $M$ at $x$, call it $T_x$, is the null space of $Q^T_x$, and the normal space to $M$  at $x$, call it $N_x$, is spanned by the columns of $Q_x$. 


We wish to sample a probability measure on $M$. We consider a measure that arises naturally in physical modelling problems, which has the form 
\begin{equation}\label{rho0}
    \rho(dx) = Z^{-1} f(x)\prod_{i=1}^m \delta(q_i(x))dx.
\end{equation}
Here $\delta$ is a Dirac delta-function, $f:\R^n\to \R$ is integrable in that sense that $\int_{\R^n} |f(x)|\prod_{i=1}^m \delta(q_i(x))dx < \infty$, and $Z$ is a normalization constant, which  is typically not known. 

This measure arises from the Boltzmann distribution when there are stiff forces that constrain a system to a neighbourhood of $M$. 
To see why, consider a contribution to a system's internal energy of the form  $\sum_{i=1}^m\half kq_i^2(x)$, where $k$ is a parameter. This gives a contribution to the partition function of the form  $e^{- \sum_{i=1}^m\half kq_i^2(x)/k_BT}$, where $T$ is temperature and $k_B$ is Boltzmann's constant. If either $k\to \infty$ (stiff forces) or equivalently $T\to 0$ (low temperature), this contribution  approaches a multiple of $\prod_{i=1}^m \delta(q_i(x))$ \cite{Fatkullin:2010vs,HolmesCerfon:2013jw}. 
For example, if particles $x_1,x_2\in \R^d$ are bound by a spring with rest length $l>0$ and Hookean energy $\half k (|x_1-x_2|-l)^2$, then as $k\to \infty$ the Boltzmann distribution approaches a multiple of $\delta(|x_1-x_2|-l)$. The stiff spring now acts as a distance constraint $|x_1-x_2|-l=0$. 

This distance constraint is however a \emph{soft} constraint -- the measure it induces depends on the particular function (or particular physical forces) used to impose the constraint, via this function's appearance in the argument of the $\delta$-function. This measure is notably different from the \emph{hard} distance constraint one would obtain by simply restricting the Boltzmann distribution to the manifold $M$ \cite{Fixman1978a,vanKampen:1981ia,Hinch:1994ca}, which gives rise to a restricted Boltzmann distribution proportional to the surface measure on $M$.

Our sampling algorithm is most easily formulated in terms of the natural surface measure (Hausdorff measure) on $M$, which we call $\mu$. 
We may convert \eqref{rho0} to a density with respect to $\mu$ using the coarea formula \cite{Morgan2016}, which is written formally as $\delta(q_i(x))dx = |Q_x|^{-1}\mu(dx)$. Here $|Q_x|$ is the pseudodeterminant of $Q_x$, equal to the product of its singular values or equivalently to $|Q^T_xQ_x|^{1/2}$, where $|\cdot|$ is the usual determinant of a square matrix. Therefore, the measure \eqref{rho0} we wish to sample, written in terms of the surface measure, is 
\begin{equation}\label{rho1}
\rho(dx) = Z^{-1}f(x)|Q_x|^{-1} \mu(dx).
\end{equation}
If one is interested in ``hard'' constraints, which don't arise from delta-functions but rather are directly imposed by restriction, then one can omit the factor $|Q_x|^{-1}$ in the density above.

\bigskip

We will sample \eqref{rho1} using a Markov Chain Monte Carlo algorithm that was proposed in \cite{Zappa:2018jy}. The algorithm generates a Markov chain $X_0,X_1,\ldots$ whose stationary distribution is \eqref{rho1}. 
Here is a high-level overview of the algorithm.  

Suppose $X_k=x\in M$. Repeat the following steps. 
\begin{enumerate}
     \item \textbf{Generate a random step in the tangent space, $v_x\in T_x$}. Let the density of this step with respect to the Euclidean measure on $T_x$ be $p_x(v_x)$. 
     \item \textbf{Project to manifold: solve $q(y)=0$ for the proposal  $y\in M$.} We require $y=x+v_x+w_x$, where $w_x\in N_x$: the projection is normal to the tangent space at the original point $x$. 
          If the numerical solver fails to find a solution, reject the proposal, and set $X_{k+1}=x$; repeat from Step 1. 
    \item \textbf{Solve for the reverse tangent step $v_y\in T_y$.} For the reverse move, from $y\to x$, we require $x = y+v_y+w_y$ where $w_y\in N_y$.
    \item \textbf{Metropolis-Hastings rejection step.} The acceptance probability is given by 
    \begin{equation}\label{acc}
        a_{xy} = \min\left( \frac{p_y(v_y)f(y)|Q_y|^{-1}}{p_x(v_x)f(x)|Q_x|^{-1}}, 1\right).
    \end{equation}
    One can omit the factors $|Q_x|^{-1},|Q_y|^{-1}$, if one is interested in hard constraints. 
    Use the acceptance probability above in a Metropolis-Hastings rejection step: 
    let $U\sim \mbox{Unif}([0,1])$. If $U > a_{xy}$, reject the move, and set $X_{k+1} = x$; repeat from Step 1. 
    \item \textbf{Reverse check: determine whether the reverse move is feasible.} Attempt to solve for $w'\in N_y$ such that $x=y+v_y+w'$. If the solver fails, or if it finds a solution such that $x' = y+v_y+w' \neq x$, then reject the proposal, and set $X_{k+1}=x$; repeat from Step 1. 
    \item \textbf{Accept proposal.} If we made it here, the proposal should be accepted: set $X_{k+1} = y$, and copy over Jacobian and Cholesky factorization, $Q_x\gets Q_y, L_x\gets L_y$. Repeat from Step 1. 
\end{enumerate}
The stationary distribution of this chain was shown in \cite{Zappa:2018jy} to be \eqref{rho1}.

\section{Numerical implementation of the algorithm}\label{sec:alg}

This section describes an efficient numerical implementation of the algorithm above to sample from \eqref{rho1}. 
There are two primary numerical challenges. One is in Step 2, the projection step, which requires solving a nonlinear system of equations. The second is in Step 5, evaluating the acceptance probability, which requires computing the pseudodeterminants $|Q_x|,|Q_y|$. We will describe a method  which efficiently addresses both of these issues. The method will require only a single matrix factorization at each iteration of the algorithm. This matrix factorization will turn out to be the primary contribution to the computational time in high dimensions. 

The matrix factorization required is the Cholesky decomposition  of $Q^T_xQ_x$. Let this decomposition be
\begin{equation}\label{cholesky}
    L_xL_x^T = Q^T_xQ_x, 
\end{equation}
where $L_x$ is a lower triangular matrix. This decomposition exists because $Q^T_xQ_x$ is positive definite, since its null space is trivial by assumption \eqref{assumption}. 

Here is how each of the steps 1-5 are implemented, in turn. Pseudocode of the algorithm is shown in Algorithm \ref{alg:pseudocode} and \ref{alg:project1}.

\algdef{SN}[propose]{StartPropose}{EndPropose}{\textbf{Choose a random step $v_x$ in tangent space:}}{}
\algdef{SN}[project]{StartProject}{EndProject}{\textbf{Project to manifold: solve for proposal $y\in M$}}{}
\algdef{SN}[revtan]{StartRevtan}{EndRevtan}{\textbf{Find reverse tangent step $v_y$:}}{}
\algdef{SN}[MH]{StartMH}{EndMH}{\textbf{Metropolis-Hastings Step:}}{}
\algdef{SN}[reverse]{StartReverse}{EndReverse}{\textbf{Reverse Projection:}}{}
\algdef{SN}[accept]{StartAccept}{EndAccept}{\textbf{Accept Proposal:}}{}

\begin{algorithm}[H]
\caption{ManifoldSampler: Given $x=X_k$, generate the next point $X_{k+1}$ in the Markov chain}
  \label{alg:pseudocode}
\begin{algorithmic}[1]
\State {Parameters:} $\sigma, \texttt{xtol}$ 
\Procedure{ManifoldSampler}{$x=X_n,Q_x,L_x$}  
 \StartPropose 
   \State Generate $\xi_i \sim N(0,1)$ for $i=1,\ldots,n$  \Comment{$\xi=(\xi_i)_{i=1}^n$ is vector of i.i.d. normals}
   \State Solve $L_xL_x^Tz = Q_x^T\xi$ for $z$ \Comment{use forward-backward substitution}
   \State $v_x\gets \sigma\cdot (\xi - Q_x z)$ \Comment{$v_x$ is isotropic Gaussian in tangent space}
    \EndPropose 
    \StartProject
   \State $y,\texttt{newtonflag} \gets$ \Call{Project}{$x+v_x,Q_x,L_x$} 
       \Comment{See Algorithm \ref{alg:project1}}
    \If{\texttt{newtonflag} == \texttt{fail}} \Comment{Projection failed to converge}
        \State Reject proposal: $X_{k+1}=X_k$. Return. 
    \EndIf
    \EndProject 
    \StartRevtan
        \State $Q_y \gets \Call{Jacobian}{q,y}$ \Comment{Jacobian $\grad q(y)^T$ provided by user}
        \State $L_y \gets \Call{Cholesky}{Q_y^TQ_y}$  \Comment{Cholesky decomposition provided by user/package}
        \State $r\gets x-y$
        \State Solve $L_yL_y^T z = Q_y^T r$ for $z$
        \State $v_y\gets r - Q_yz$
    \EndRevtan 
    \StartMH
        \State $\texttt{vdiff} \gets (|v_y|^2 - |v_x|^2) / (2\sigma^2)$
        \State $\texttt{udiff} \gets \log(f)(y) - \log(f)(x)$ \Comment{Function $\log(f)$  provided by user}
        \State $\texttt{qdet} \gets \prod_i l_i$,\quad  $l_i  =(L_x)_{ii}/(L_y)_{ii}$ \Comment{Set \texttt{qdet}=1 if you have hard constraints}
        \State $a\gets \exp(-\texttt{vdiff})*\exp(-\texttt{udiff})*\texttt{qdet}$ \Comment{Acceptance probability}
        \If{$U\sim \text{Unif}([0,1]) > a$} 
              \State Reject proposal: $X_{k+1}=X_k$. Return. 
        \EndIf  
    \EndMH 
    \StartReverse
        \State $x',\texttt{newtonflag} \gets$ \Call{Project}{$y+v_y,Q_y,L_y,v_y$}
        \If{$\texttt{newtonflag} == \texttt{fail}$}   \Comment{Reverse projection failed to converge}
           \State Reject proposal: $X_{k+1}=X_k$. Return. 
        \EndIf
        \If{$|x'-x| > \texttt{xtol}$}  \Comment{Projection converged, but to wrong point}
          \State Reject proposal: $X_{k+1}=X_k$. Return. 
       \EndIf
    \EndReverse 
    \StartAccept
        \State $x\gets y$, $Q_x\gets Q_y$, $L_x\gets L_y$  \Comment{Copy matrices/factorizations by exchanging pointers}
    \EndAccept
\EndProcedure
\State \textbf{end procedure}
\end{algorithmic}
\end{algorithm}

\begin{algorithm}[h]
\caption{Project to manifold (symmetric Newton): Solve $q(y)=0$ s.t. $y=z+Qa$}
  \label{alg:project1}
\begin{algorithmic}[1]
\State {Parameters:} $\texttt{tol},\eta,\texttt{MaxIter}$ 
\Procedure{Project[SymmetricNewton]}{$z,Q,L$} 
    \State $a \gets 0$-vector \Comment{Initial guess (must be deterministic)}
    \State $y\gets z + Q*a$, $\texttt{qerr}\gets \infty$ \Comment{Initialization}
    \For{$\texttt{niter}=0:\texttt{MaxIter}-1$}
    \State $\texttt{qerr0}=\texttt{qerr}$  \Comment{Save error from previous step}
        \State $\texttt{qval}\gets q(y)$  \Comment{Evaluate constraints}
        \State $\texttt{qerr} \gets |q(y)|_\infty$  \Comment{Could use other norms}
        \If{\texttt{qerr} < \texttt{tol}}    \Comment{We found a solution}
        \State return $y$, \texttt{Success} 
        \EndIf
        \If{$\texttt{qerr} > \eta*\texttt{qerr0}$} \Comment{Error isn't decreasing fast enough; no solution}
        \State return $y$, \texttt{Fail} 
        \EndIf
        \State Solve $LL^t\delta a = \texttt{qval}$ for $\delta a$ \Comment{Forward-backward substitution}
        \State $a \gets a+\delta a$, $y\gets z+Q*a$ \Comment{Next guess}
    \EndFor
    \State\textbf{end for}
    \State return $y$, \texttt{Fail} \Comment{We didn't find a solution to sufficient accuracy}
\EndProcedure
\State \textbf{end procedure}
\end{algorithmic}
\end{algorithm}

\begin{algorithm}[h]
\caption{Project to manifold (traditional Newton): Solve $q(y)=0$ s.t. $y=z+Qa$}
  \label{alg:project2}
\begin{algorithmic}[1]
\State {Parameters:} $\texttt{tol},\eta,\texttt{MaxIter}$ 
\Procedure{Project[TraditionalNewton]}{$z,Q$} 
    \State $a \gets 0$-vector \Comment{Initial guess (must be deterministic)}
    \State $y\gets z + Q*a$, $\texttt{qerr}\gets \infty$ \Comment{Initialization}
    \For{$\texttt{niter}=0:\texttt{MaxIter}-1$}
    \State $\texttt{qerr0}=\texttt{qerr}$  \Comment{Save error from previous step}
        \State $\texttt{qval}\gets q(y)$  \Comment{Evaluate constraints}
        \State $\texttt{qerr} \gets |q(y)|_\infty$  \Comment{Could use other norms}
        \If{\texttt{qerr} < \texttt{tol}}    \Comment{We found a solution}
        \State return $y$, \texttt{Success} 
        \EndIf
        \If{$\texttt{qerr} > \eta*\texttt{qerr0}$} \Comment{Error isn't decreasing fast enough; no solution}
        \State return $y$, \texttt{Fail} 
        \EndIf
        \State $Q_y\gets \Call{Jacobian}{q,y}$
        \State $L,U\gets \Call{LU}{Q_y^TQ}$  \Comment{LU decomposition provided by user/package}
        \State Solve $LU\delta a = \texttt{qval}$ for $\delta a$ \Comment{Forward-backward substitution}
        \State $a \gets a+\delta a$, $y\gets z+Q*a$ \Comment{Next guess}
    \EndFor
    \State\textbf{end for}
    \State return $y$, \texttt{Fail} \Comment{We didn't find a solution to sufficient accuracy}
\EndProcedure
\State \textbf{end procedure}
\end{algorithmic}
\end{algorithm}


\subsection{Generate a random step in the tangent space $v_x\in T_x$}\label{sec:v}

We generate an isotropic Gaussian in the tangent space. 
Let $\xi\in \R^n$ be a vector of i.i.d. standard normals. We find $v_x$ by projecting $\xi$ orthogonally onto the tangent space, and then scaling by a parameter $\sigma>0$ which controls the overall size of the step. The orthogonal projection onto the 
tangent space is $(I-Q_x(Q_xQ_x^T)^{-1}Q_x^T)\xi$.
We implement this projection without matrix inversion,  by solving 
\begin{equation}\label{proj1}
    L_xL_x^Tz = Q^T_x \xi \qquad \text{for $z$}, 
\end{equation}
using the Cholesky decomposition \eqref{cholesky} to solve the system of equations by forward-backward substitution.
Then we set 
\begin{equation}\label{proj2}
    v_x = \sigma(\xi - Q_x z). 
\end{equation}
The density of this tangent space proposal is 
\begin{equation}\label{pv}
    p_x(v_x) = \frac{1}{(2\pi\sigma^2)^{d/2}}e^{-\frac{|v_x|^2}{2\sigma^2}}.
\end{equation}

In \cite{Zappa:2018jy} it was suggested to choose $v_x$ by first computing an orthonormal basis  of the tangent space $T_x$ using the QR decomposition of $Q_x$ (from the last $d$ columns of the Q-matrix), storing the basis in the columns of a matrix $E_x$, and then setting $v_x=\sigma E_x\xi'$ where $\xi'\in \R^d$ is a vector of i.i.d. standard normals. 
In high dimensions, a similar approach could be implemented using a sparse QR decomposition of $Q_x$, which represents the Q-matrix as a product of Householder transformations or Givens rotations \cite{Davis2006}, thereby producing a basis of the normal space $N_x$ which can be used for projection. However, we expect such an approach to be less efficient than the one proposed above, partly because the QR decomposition is more expensive to compute than the Cholesky decomposition, and partly because  we will require the Cholesky decomposition in later steps of the algorithm in places where it cannot be substituted with the QR decomposition.

We remark that constrained molecular dynamics simulations (e.g. \cite{Barth:1995vi}) generate an initial proposal in the full, ambient space, equivalent to our random vector $\xi$, and then project to $M$ to find $y=x+\sigma \xi+w$, thereby avoiding the need to calculate the tangent step. 
It is not possible to Metropolize such a proposal without additional processing, because for a given $x,y$, there is a positive-dimensional space of possible $\xi,w$, hence, we cannot determine the probability density of $\xi$ which is required to evaluate the acceptance probability.

\subsection{Project to manifold: Solve for the proposal  $y\in M$ }\label{sec:project}

The next step is to solve for $y$ such that $q(y) = 0$, with  $y=x+v_x+w_x$ and $w_x\in N_x$ is the vector to be determined. Observe that we can write any $w\in N_x$ as $w = \sum_{i=1}^m a_i\grad q_i(x)$ for a unique choice of $a = (a_1,\ldots,a_m)^T \in \R^m$, by assumption \eqref{assumption}. Therefore,  we must solve the system of equations
\begin{equation}\label{system}
      q_i\left(x+v_x+\sum_{i=1}^m a_i\grad q_i(x)\right) = 0, \qquad i=1,\ldots, m,
\end{equation}
for the unknown $a$. 

We propose to solve this system using a quasi-Newton method that replaces the exact Jacobian with an approximate Jacobian for which we already have a factorization; we call this \textbf{symmetric Newton}, following \cite{Barth:1995vi}. We compare this method with Newton's method, which we refer to as \textbf{traditional Newton}.  One contribution of this paper is to demonstrate that symmetric Newton \emph{significantly} increases the overall efficiency of the algorithm compared to traditional Newton. 

Let us first describe \textbf{traditional Newton}'s method to solve \eqref{system}. 
(Pseudocode is found in Algorithm \ref{alg:project2}.)
The method generates a sequence of approximations $a\e{0}, a\e{1}, a\e{2}, \ldots$ with corresponding estimates of $y$ computed as $y\e{n} = x+v_x+Q_xa\e{n}$. 
The Jacobian of \eqref{system} is $Q_{y\e{n}}^TQ_x$, so the iterates are updated as 
 $a\e{n+1} = a\e{n} + \delta a\e{n}$, where $\delta a\e{n}$ solves
\begin{equation}\label{QyQx}
    Q_{y\e{n}}^TQ_x \delta a\e{n} = -q(y\e{n}). 
\end{equation}
We solve the above linear system of equations using an LU decomposition of $Q_{y\e{n}}^TQ_x$, and initial condition $a\e{0} = 0$, the zero vector. 
The method iterates until a stopping criterion is satisfied; we discuss the stopping criterion at the end of this section. Notice that at each iteration,  we must evaluate  $Q_{y\e{n}}$ and compute an LU decomposition of $Q_{y\e{n}}^TQ_x$, both of which are expensive operations in high dimensions. 

Our second method, \textbf{symmetric Newton}, replaces the exact Jacobian $Q_{y\e{n}}^TQ_x$ with an approximation $Q_{x}^TQ_x$. (Pseudocode is found in Algorithm \ref{alg:project1}.) We already have its Cholesky factorization \eqref{cholesky} which we used to generate a tangent step in Section \ref{sec:v}, so there is no further Jacobian evaluation nor matrix factorization required.  
Therefore, symmetric Newton computes updates by solving 
\begin{equation}\label{QxQx}
    Q_{x}^TQ_x \delta a\e{n} = -q(y\e{n}), 
\end{equation}
using the Cholesky decomposition \eqref{cholesky} to solve this linear system of equations by forward-backward substitution. 

We expect symmetric Newton to converge to the solution more slowly than traditional Newton, since it should converge linearly to the true solution whereas traditional Newton converges quadratically \cite{Nocedal:2006uv}. It could also fail to converge in some cases even when Newton's method converges, leading to an increase in rejections. However, we also expect an increase in efficiency gained by avoiding a Jacobian evaluation and matrix factorization at each iteration of the solver. 
It is not immediately clear how this tradeoff will play out; we will determine this via examples. 

We turn to discussing the \textbf{stopping criterion}, which is the same for both Newton methods. The stopping criterion is important because it is inefficient to spend a lot of computational effort searching for a solution, when we could instead give up easily and try again in a different random direction. 
Furthermore, there is sometimes no solution to \eqref{system}, and indeed the sampler can be more efficient when operated in a regime where there is frequently no solution \cite{Lelievre:2019be}; it is important not to waste time searching for a solution that doesn't exist. 

We have found the following stopping criterion to be effective. Terminate Newton's method  when any one of the following is satisfied: 
\begin{enumerate}[(i),nosep]
    \item $\texttt{err}_n < \texttt{tol}$, where $\texttt{tol}$ is a tolerance parameter determining how close we are to the manifold, and $\texttt{err}_n = |q(y\e{n})|_\infty = \max_i|q_i(y\e{n})|$ is a measure of error. When this happens we return a \emph{success}: we have found a solution sufficiently close to $M$. 
    \item $\texttt{err}_n > \eta \texttt{err}_{n-1}$, where $\eta$ is a parameter governing how much the error must shrink at each step (typically $\eta<1$). If this criterion is satisfied, we return a \emph{fail}. 
    \item $n \geq \texttt{MaxIter}$, where $\texttt{MaxIter}$ is a parameter controlling the maximum number of iterations. If this criterion is satisfied, we return a \emph{fail}. Because of the previous stopping criterion, we rarely (if ever) reach this step. 
\end{enumerate}

\bigskip


We explored other stopping criteria that would allow us to spend more time looking for a solution. For example, we explored a method where if case (ii) occurs, then instead of stopping, we tried shrinking the Newton step size by half, i.e. we set $\delta a\e{n} \to \delta a\e{n}/2$, and re-computed the error, and re-evaluated this stopping criterion; repeating the shrinking step several times if necessary. We found this increased the number of acceptances by a negligible amount, so we dropped this extra method.


\subsection{Solve for the reverse tangent step $v_y\in T_y$}\label{sec:vp}

The reverse move proposes $x=y+v_y+w_y$ with $v_y\in T_y$ and $w_y\in N_y$. Since, $v_y,w_y$ live in orthogonal subspaces, they may each be determined by projecting $x-y$ orthogonally onto the appropriate subspace. Therefore, we find $v_y$ using the same steps as in \eqref{proj1}, \eqref{proj2}: 
\begin{equation}\label{reversev}
    \text{Solve } \quad L_yL_y^Tz' = Q_y^T(x-y) \quad \text{ for $z'$;} \qquad
    \text{set} \quad v_y = (x-y)-Q_yz'. 
\end{equation}
To do this we must first evaluate $Q_y$ and compute the Cholesky decomposition \eqref{cholesky} of $Q_yQ_y^T$. 
In practice, at each iteration of the Monte-Carlo algorithm, \emph{this is the only step where a Jacobian evaluation and a Cholesky decomposition is computed} -- we have the Cholesky decomposition of $Q_xQ_x^T$ because we computed it when $x$ was first proposed.

\subsection{Metropolis-Hastings rejection step}

We test whether the acceptance criterion is satisfied \emph{before} we perform the reverse check (which is also necessary for the acceptance criterion to hold), because it is more efficient in this order: if we reject a move because of the Metropolis criterion, which compares $f,v_{(\cdot)},|Q_{(\cdot)}|^{-1}$, then we do not need to proceed to a reverse projection, which is computationally much more expensive.  

Evaluating the acceptance probability in high dimensions requires a few remarks on implementation. The main challenge is that the terms in \eqref{acc} become increasingly tiny as the dimension is increased, resulting in numerical underflow. The solution is generally to work with the logarithms of the quantities involved. 
That is, to compute  $f(y)/f(x)$ we provide a function which returns $\log f(x)$, then exponentiate the difference $\log f(y) - \log f(x)$. 
For the ratio  $p_y(v_y) / p_x(v_x) = e^{-|v_y|^2/2\sigma^2} / e^{-|v_x|^2/2\sigma^2}$ (see \eqref{pv}) we first compute $-(|v_y|^2 - |v_x|^2)/2\sigma^2$, and then exponentiate. 

The final term to evaluate is $|Q_y|^{-1}/|Q_x|^{-1}$. 
These pseudodeterminants may be expressed in terms of the Cholesky factorizations \eqref{cholesky} as
\begin{equation}
    |Q_x| = |Q_xQ_x^T|^{1/2} = \prod_{i=1}^m (L_x)_{ii}, \qquad
     |Q_y| = |Q_yQ_y^T|^{1/2} = \prod_{i=1}^m (L_y)_{ii}.
\end{equation}
In high-dimensional problems where these determinants result in numerical underflow, we evaluate the contribution to the acceptance probability by computing the ratio of individual diagonal elements of the Cholesky matrices, and then multiplying these ratios together: 
\begin{equation}\label{det}
r_i = \frac{(L_x)_{ii}}{(L_y)_{ii}}, \qquad \quad
    \frac{|Q_y|^{-1}}{|Q_x|^{-1}} = \prod_{i=1}^m r_i.
\end{equation}
This is effective as the corresponding diagonal elements $(L_x)_{ii}$, $(L_y)_{ii}$, tend to be similar to each other. 

We could alternatively compute $\log|Q_{(\cdot)}| = \sum_i\log (L_{(\cdot)})_{ii}$ and then compute the ratio of determinants as $|Q_y|^{-1}/|Q_x|^{-1} = \exp\left(\log |Q_x| - \log |Q_y|\right)$. We verified that this gives the same ratio of determinants and avoids numerical underflow, however we prefer \eqref{det} as it avoids computing logs and exponentials.

\subsection{Reverse check: determine whether the reverse move is feasible}   

The nonlinear solver in Section \ref{sec:project} does not always produce a solution to \eqref{system}, even if one exists. Therefore, it is possible that the solver produces a proposal $y\in M$, but then when asked to solve for $x'\in M$ such that $x' = y+v_y+w'$, it fails to find a solution. If this happens we must reject the proposal, to satisfy detailed balance. (See \cite{Zappa:2018jy} for an enhanced discussion of this issue.) 

We check this possibility by attempting to solve \eqref{system} for $x'$, starting from $y+v_y$, using exactly the same solver as for the forward move: either traditional Newton, or symmetric Newton. If the solver fails, or if it finds a solution $x'\neq x$ (in practice, such that $|x'-x| > \texttt{xtol}$, where \texttt{xtol} is a tolerance parameter), we reject the proposal. 

This additional nonlinear solve increases the computational time. 
Whether or not it this additional time is worth it, depends on the problem. Ref. \cite{Lelievre:2019be} found that in their Hamiltonian sampler, for the low-dimensional examples they considered, their sampler was more efficient when operated in a regime where there was a significant fraction (up to 15\%) of moves rejected during this step. In 3 out orfour 4 examples, we find a negligible  percentage  of moves are rejected during this step (see Section \ref{sec:reject}), except occasionally in the lowest-dimensional versions of each example.  Therefore, in many cases, this extra projection check could be dropped, at the expense of introducing a slight bias in the measure being sampled.

\section{Numerical Examples and Results}\label{sec:results}

We apply our algorithm to four different examples, described in Section \ref{sec:examples}. We aim to demonstrate the algorithm is effective in systems with $O(1000)$ variables, in the sense that it allows generating $O(10^6)$ points or more on reasonable timescales, e.g. no more than a timescale of days. We make no claims about the \emph{convergence} of a Monte Carlo method on these examples (which would require analyzing for example the correlation time of statistics of interest), as this depends sensitively on each example and introduces considerations beyond the numerical ones considered in this paper. We did verify that the sampler produces points with the correct distribution, on some of the examples considered in \cite{Zappa:2018jy,HolmesCerfon:2020dk} for which a nontrivial analytic marginal distribution is known.

We further wish to compare the two projection methods, and show that the symmetric Newton projection method significantly increases the efficiency of the sampler compared to traditional Newton, by a factor of 3-23 depending on the example. In some examples, this makes an otherwise intractable problem tractable. We show that a primary reason for the increase in efficiency is it reduces the number of matrix factorizations.  


We make a few remarks on implementation. 
We implement our algorithm in C++. We use Eigen for linear algebra operations, an open-source library for linear algebra computations, that is comparable to other high-performance linear algebra libraries. 
We consider problems where $Q_x$ is sparse. If it isn't, then constraint-based methods will be inefficient. We use AMD ordering to compute a Cholesky desompotion, and COLAMD ordering for LU decompositions, which we found empirically to be slightly faster than AMD (Eigen does not have COLAMD ordering for Cholesky decompositions). 
The algorithm is compiled and run on a 2.2GHz Intel Xeon processor. 
These implementation choices will affect the timings reported, but are not expected to affect their qualitative behaviour; furthermore we provide heuristics for estimating timings for other implementations. 

We chose parameters \texttt{tol} = $10^{-5}$, $\eta=0.95$, \texttt{MaxIter} = 100 for the projection step, and $\texttt{xtol}=\texttt{tol}*n*10$. The maximum number of iterations was rarely (if ever) reached, because of the other termination criteria.

Code which implements all of the examples below and several other simpler ones can be found at \cite{gitcode}. 

\subsection{Examples}\label{sec:examples}

We test our algorithm on four examples. Each has a parameter, $n$, controlling the size of the problem; in most examples $n$ represents a number of particles or vertices (this is different from how we used the variable $n$ in the previous sections). The number of variables $n_{\rm vars}$ and the number of constraints $m$ are each proportional to $n$. The examples explore different regimes for the fraction of constraints $m/n_{\rm vars}$, both small to large, and different patterns for the constraints, leading to different sparsity patterns in $Q_x$. In all but one example we set $f(x)=1$ to focus purely on the geometrical aspects of the sampler. Problems where $f$ is significant and dominate the computational time were explored in \cite{Barth:1995vi}.

\begin{figure}
\includegraphics[width=0.4\textwidth]{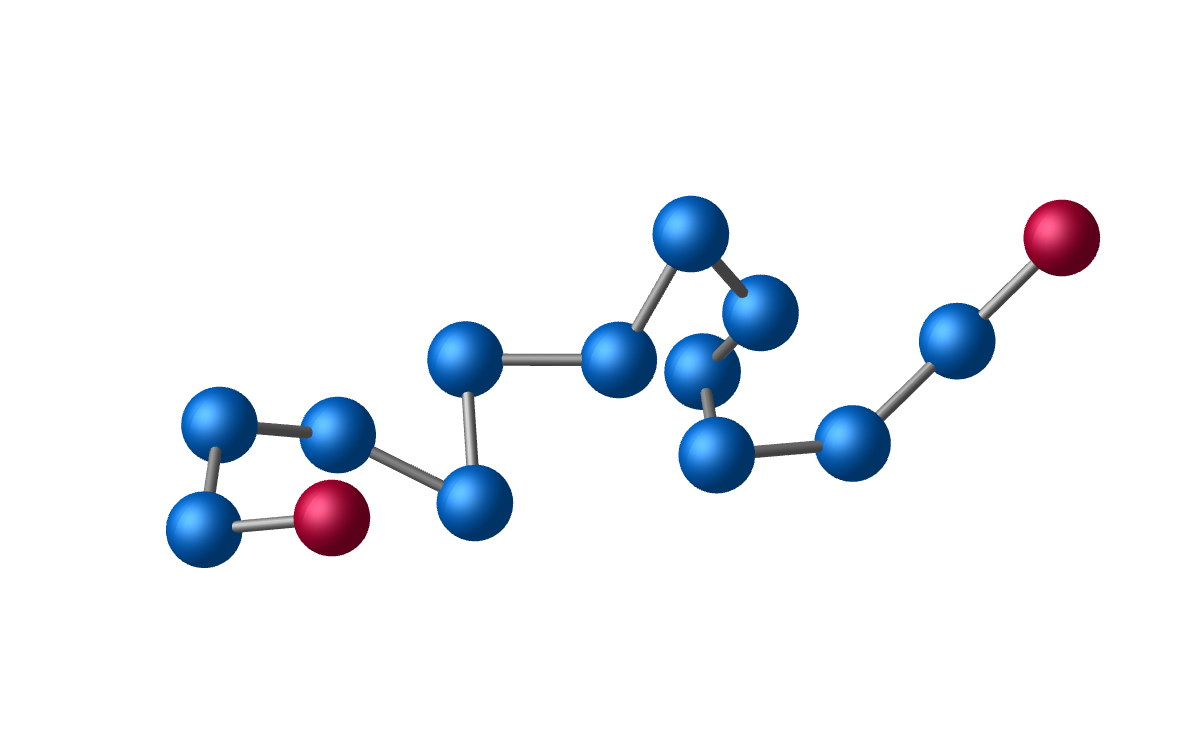}
\includegraphics[width=0.3\textwidth]{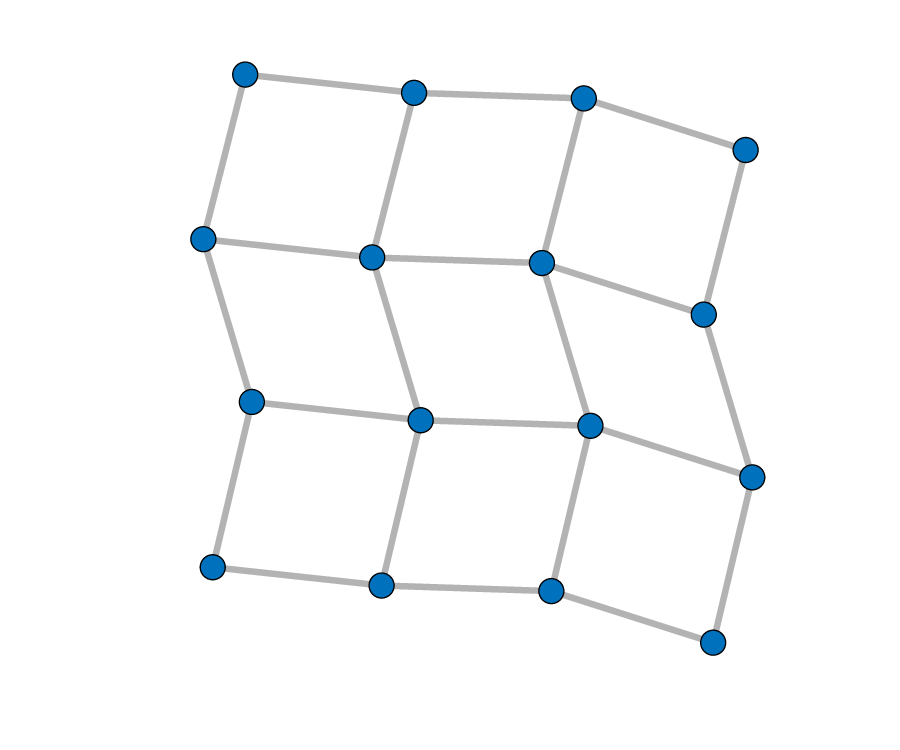}
\includegraphics[width=0.25\textwidth]{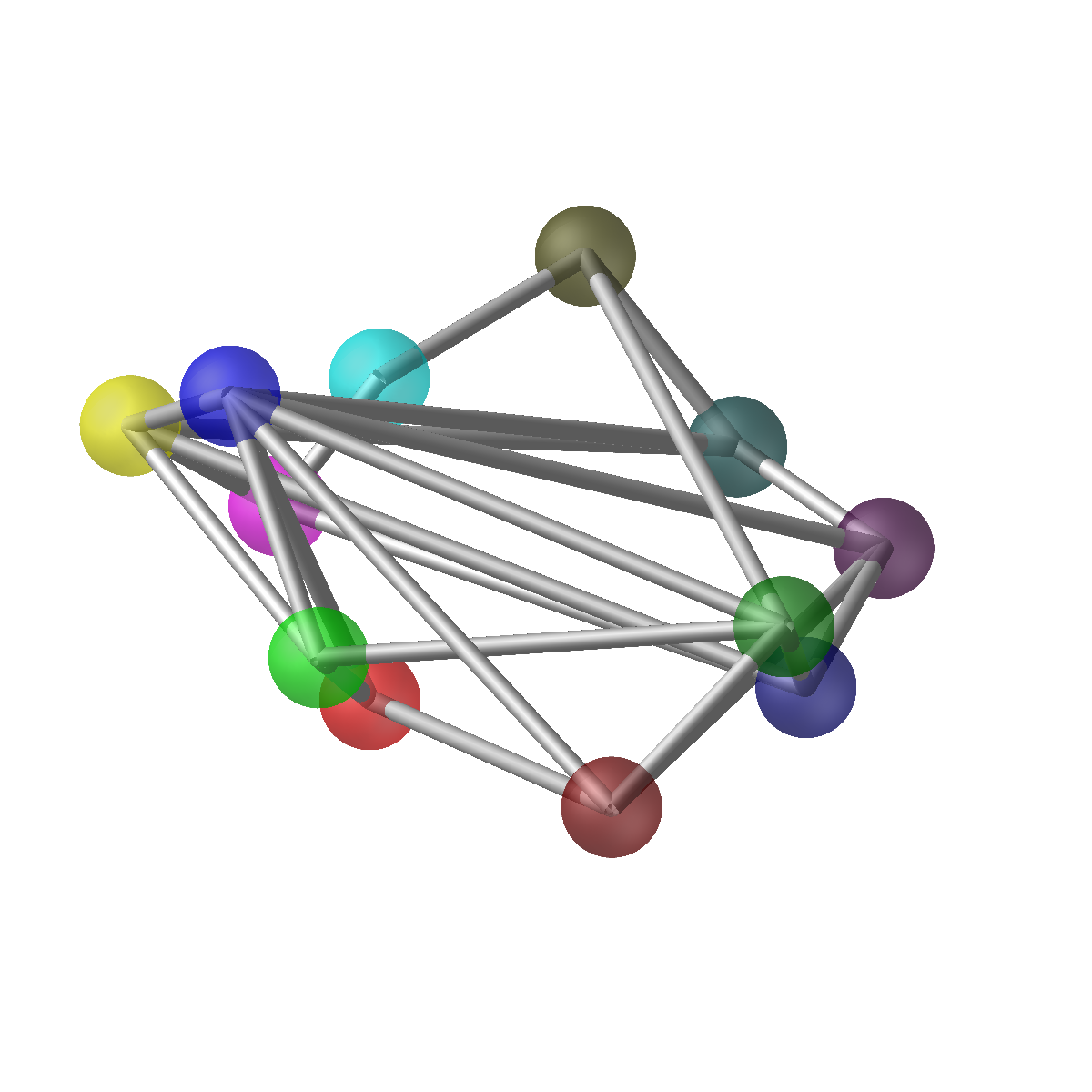}
\caption{Illustrations of selected examples. Left: Polymer (Example \ref{ex:polymer}) with $n=12$ particles, with fixed endpoints shown in red. Middle: Lattice (Example \ref{ex:lattice}) with $n=16$ particles. Right: Ngon with extra random edges (Example \ref{ex:polygon}) with $n=12$ particles. 
}\label{fig:images}
\end{figure}

\subsubsection{Polymer with fixed endpoints}\label{ex:polymer}

We consider a three-dimensional polymer with $n$ vertices, bound by bars of length 1, whose endpoints are bound with bars of length 1 to fixed vertices at $a_0=(0,0,0)$ and $a_1=(n/2,0,0)$; see Figure \ref{fig:images}.  This models a polymer of length $n+2$ with fixed endpoints, whose end-to-end length is fixed at roughly half the total length of the polymer.  We remark that a polymer without endpoint constraints is straightforward to parameterize and sample directly, while adding the endpoint constraints makes sampling significantly more challenging and requires specialized methods \cite{Oyarzun:2018hz}. 
We have
\[
   n_{\rm vars}  =  3n, \qquad 
   m = n+1.
\]
In this example the number of constraints is reasonably small, $m/n_{\rm vars} \approx 1/3$. For point particles in three dimensions, this is the smallest scaling for the number of distances constraints that allows for a connected set of vertices.

The system is described by the positions of the vertices, $x = (x_1,\ldots,x_n)^T\in \R^{3n}$ where $x_i\in \R^3$ is the position of the $i$th vertex. 
The bars are implemented with constraints of the form 
\begin{equation}
    q_k(x) = |x_k-x_{k+1}|^2 - 1, \qquad k=1,\ldots,m-1,
\end{equation}
and for the endpoints we have $q_m(x) = |x_1-a_0|^2-1$, $q_{m+1}(x) = |x_n - a_1|^2 - 1$. 
We set $f(x) = 1$. 

We remark that although it might be more physical to sample a constraint of the form $q_i(x) = |x_i-x_{i+1}| - 1$, since this is the constraint that would arise if the bar were a stiff spring, one can verify that $|Q_x|$ varies by a constant between the two cases, so it doesn't alter the acceptance probability. Therefore, we choose the constraint whose gradient is less expensive to evaluate.

\subsubsection{Lattice}\label{ex:lattice}

We consider a two-dimensional square lattice with $n$ vertices bound by bars of unit length, modelling a crystalline material, see Figure \ref{fig:images}. We must choose $n=s^2$ where $s$ in an integer, representing the number of vertices on each side of the lattice.
We have 
\[
   n_{\rm vars}  =  2n, \qquad 
   m = 2n-2\sqrt{n}.
\]
This example has close to the maximum number of constraints, $m/n_{\rm vars} =1-1/\sqrt{n}\approx 1$.

The system is described by $x = (x_1,\ldots,x_n)^T\in \R^{2n}$ where $x_i\in \R^2$ is the position of the $i$th vertex. The constraints have the form 
\begin{equation}
    q_k(x) = |x_{i_k}-x_{j_k}|^2 - 1, 
\end{equation}
where $k$ indexes the constraints and where $(i_k,j_k)$ give the vertices associated with the $k$th edge. 

We found that without imposing an additional energy function on the lattice, it collapsed in our simulations to a configuration where one line of vertices was nearly on top of another. The sampler was unable to successfully make a proposal $y$ from this configuration, so it remained in that configuration thereafter. Such a configuration is geometrically ``singular'', with $|Q_x|\approx 0$, violating our assumption \eqref{assumption}. 

Therefore we impose an energy on the lattice to prevent it from collapsing. This is the only example we consider with a nontrivial energy, so this energy also serves to explore a regime where the energy function is nontrivial. We found the following function to be effective, allowing the lattice to deform slightly yet not collapse:
\begin{equation}
f(x) = \exp(-k \sum_i (d_i- \sqrt{n}/\sqrt{2} )^2).
\end{equation}
Here $d_i$ is the length of a diagonal for a small unit block and $k=5$ is a parameter. The sum indexed by $i$ runs over all the diagonals in both directions of the lattice. 

\subsubsection{Orthonormal matrix}\label{ex:matrix}

We consider the special orthogonal group of $n=s\times s$ orthonormal matrices with positive determinant, $SO(s) = \{A\in \R^{s\times s}: A^TA= I, |A|=1\}$. Each matrix can be reshaped into a vector with $n=s^2$ entries. The set $SO(s)$ is characterized by a condition on the dot products of its rows, which form our constraints:
\[
A_{i,\cdot}^TA_{i,\cdot} -1 = 0, \qquad
A_{i,\cdot}^TA_{j,\cdot}  = 0 \quad (i\neq j).
\]
Here $A_{i,\cdot}$ is the $i$th row of $A$. 
We have 
\[
   n_{\rm vars}  =  n, \qquad 
   m = \frac{n+\sqrt{n}}{2}.
\]
Here the number of constraints is $m/n_{\rm vars} \approx 1/2$.  

We wish to sample points in proportion to the natural surface measure (the Haar measure) on this space. It can be shown that $|Q_x|$ is constant on $SO(s)$ for the constraints as written above, so we keep this factor in our sampling as it makes no difference to the acceptance probability. We also note that one can show that $|v_x|=|v_y|$ for all $x,y\in SO(s)$. Therefore, with $f(x) = 1$, the acceptance probability should be identically $1$, and the only way to reject a proposal should be if the projection step fails to converge. 

We remark that our sampler is not the best way to sample from the uniform distribution on $SO(s)$ \cite{Mezzadri2007}. We include this example merely to test our sampler with an additional type of constraint structure.

\subsubsection{N-gon with random vertices and edges }\label{ex:polygon}

Our next example considers random graphs, chosen to have a fairly large number of constraints, and such that there is no structure in the sparsity pattern in $Q_x$. We consider $n$ vertices in three-dimensional space, which are initially equally spaced along a circle in the $xy$-plane such that the distance between neighbouring points on the circle is 1. The vertices are then connected by  $n$ edges to form a polygon, and then an additional $n$ edges are chosen randomly from the remaining pairs. All edges represent distance constraints. The vertices are then perturbed: the vertical ($z$) coordinates are set to be independent Gaussian random variables with standard deviation 0.5, while the horizontal radial coordinate is multiplied by an independent random variable that is uniform on $[0.6,1]$. (These perturbations allow the graph to have some flexibility; the particular choices of perturbations are not important.) The lengths of all edges are then computed from this configuration. 
See Figure \ref{fig:images} for one such random graph. 

For this system
\[
n_{\rm vars} = 3n, \qquad m = 2n,
\]
and hence $m/n_{\rm vars} = 2/3$. 
Because each realization of the example is different, we typically consider 3 random examples for each value of $n$. We always use $f(x) = 1$.

\subsection{Comparing projection methods: Traditional Newton versus Symmetric Newton}\label{sec:timing}

We are interested in comparing the efficiency of the algorithm for each type of projection algorithm, traditional Newton and symmetric Newton. 
Each method has a parameter $\sigma$ governing the size of the tangent step, and any measure of efficiency will be sensitive to this parameter. Comparing the methods therefore requires choosing a value of $\sigma$ for each method that allows for a fair comparison. 
Ideally we should choose $\sigma$ to minimize the computational time required to generate an ``effectively independent'' sample point. However, because measuring the level of convergence of a Markov chain is computationally not straightforward \cite{Cowles1996,Qin2022}, a common practice is to choose a target average acceptance ratio $a$ and tune the step size parameter $\sigma$ to achieve this ratio. Theory on idealized problems suggests that  $a\approx 0.234$ leads to optimal correlation times in high dimensions \cite{Gelman1996,Gelman1997}.


We follow this same practice, and determine for each example and each method the value of $\sigma$ which leads to a desired average acceptance ratio $a$; call these values $\sigma\e{N}_a, \sigma\e{S}_a$ for traditional Newton and symmetric Newton respectively. For most of our tests we choose $a=0.25$; subsequent tests (Section \ref{sec:diff}) show that changing the value of $a$ doesn't alter our qualitative observations.

\begin{figure}
\includegraphics[width=\textwidth]{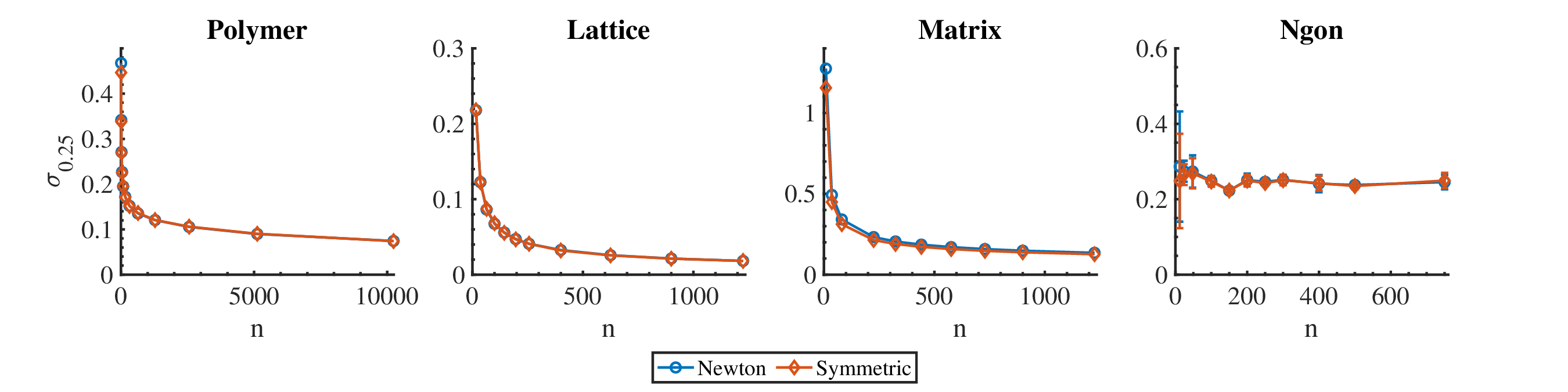}
\caption{Estimated values of $\sigma_{0.25}$, the step-size parameter that leads to average acceptance probability $a=0.25$, as a function of system size parameter $n$, for our four examples. The data for the N-gon is an average over 3 independent examples at each $n$, with $\sigma_{0.25}$ determined separately for each example, and is shown with one-standard-deviation error bars.
}\label{fig:sig}
\end{figure}

Figure \ref{fig:sig} shows the estimated values of $\sigma\e{N}_{0.25}, \sigma\e{S}_{0.25}$ for each example at different system sizes $n$. 
These were estimated using a bisection algorithm that repeats until the value of $a$ estimated using $10^5$ sampled points is within $1.25/\sqrt{10^5}$ of the target value $a$. 
 The values of $\sigma\e{N}_{0.25}, \sigma\e{S}_{0.25}$ are generally very close. A slight exception is the matrix (Example \ref{ex:matrix}), where $\sigma\e{N}_{0.25}$ is about 8-10\% larger than $\sigma\e{S}_{0.52}$ (the difference decreases slightly with $n$). We would expect this difference to be associated with a $(\sigma\e{N}_{0.25}/\sigma\e{S}_{0.25})^2\approx16-20$\% increase in overall efficiency for traditional Newton, assuming diffusive dynamics. 
As a side remark, the step sizes decrease with $n$ as $\sim n^{-0.2},n^{-0.5},n^{-0.3}$ for the Polymer, Lattice, and Matrix respectively. They remain roughly constant for the Ngon, presumably because the geometry of the example changes with dimension.

\begin{figure}
\includegraphics[width=1\textwidth]{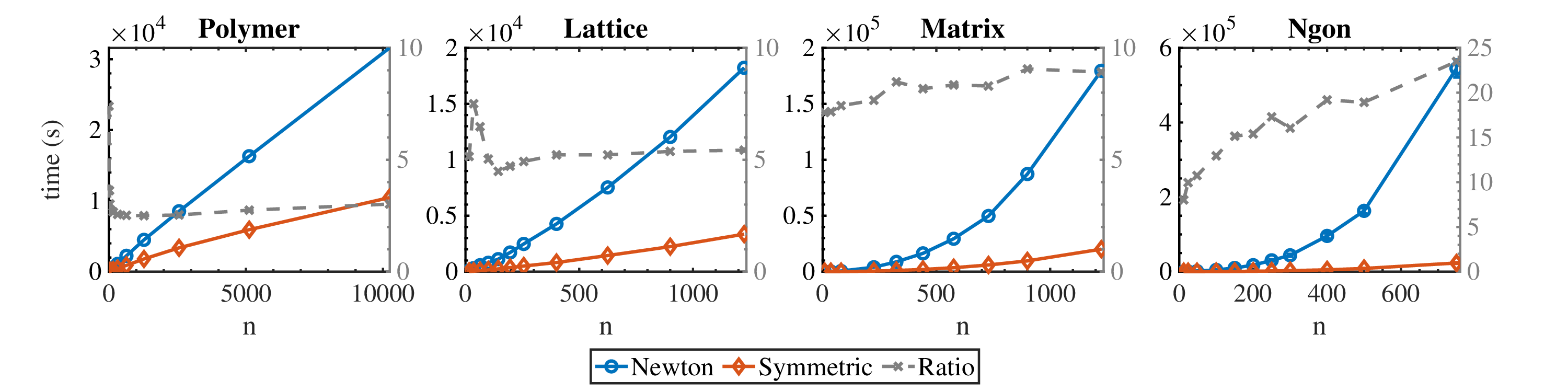}
\caption{Time (in seconds) to sample $10^6$ points for each example, using step-size parameter $\sigma_{0.25}$, for each of traditional Newton and symmetric Newton. The ratio (traditional:symmetric) is shown in grey on the right axes. Data for the Ngon is an average over 3 independent examples; the one-standard deviation error bars are smaller than the markers.
}\label{fig:time}
\end{figure}

\begin{table}
\centering
\begin{tabular}{l | cccc}
     & Polygon & Lattice & Matrix & Ngon \\\hline
  Newton & 0.98 & 1.27 & 2.24 & 2.35 \\
  Symmetric & 0.99 & 1.21 & 2.16 & 2.08 \\\hline
  LU & 0.97 & 1.36 & 2.35 & 2.48 \\
  Cholesky & 0.97 & 1.56 & 2.79 & 2.75 \\\hline
\end{tabular}
    \caption{Complexity of each method or matrix factorization, reported as the power $\alpha$ in the scaling law $\text{time}\sim O(n^\alpha)$, obtained by using least-squares to fit our empirical timing data (minus the 3 smallest values of $n$) on a log-log scale. }\label{tbl:complexity}
\end{table}

We then measured the total computational time to generate $10^6$ points for each example at several values of $n$, using the estimated step sizes $\sigma\e{N}_{0.25}, \sigma\e{S}_{0.25}$. Figure \ref{fig:time} shows that
symmetric Newton is \emph{uniformly} faster than traditional Newton, by a factor that depends on the example: for the Polymer it is roughly 2.5-3 times faster; for the Lattice it is roughly 5 times faster; for the Matrix it is roughly 7-9 times faster (though accounting for the slightly greater step size for traditional Newton, the actual speedup is about a factor of 6-8). 
The Ngon shows the greater speedup, and the strongest dependence on $n$, increasing from an average of about 8 times faster for $n=12$ to more than 23 times faster for $n=750$. 
Notably, this increase changes the order of magnitude of the computations: at $n=750$, symmetric Newton took just over $6$ \emph{hours} while traditional Newton took just over $6$ \emph{days}. 

The computational complexities of traditional and symmetric Newton are similar (Table \ref{tbl:complexity} ), and depend on the example, being smallest for the Polymer, roughly $O(n)$, and largest for the Matrix and Ngon, roughly $O(n^{2.1})$. The complexity for traditional Newton is slightly higher than for symmetric Newton. The complexities are similar to the complexities of performing sparse matrix factorizations for each example (Section \ref{sec:matrixfac}). 

We remark that we verified that the actual measured acceptance ratio $\hat{a}$ for the $10^6$ points was close to the desired one, in the range $\hat{a}\in[0.24,0.26]$. 
For the Ngon the actual acceptance ratio sometimes varied more than this. We suspect this could be because the random geometry caused the configuration space of the N-gon to contain some regions allowing for different (usually larger) step sizes, but that reaching these regions was a rare event requiring a long burn-in time. When $\hat{a}$ was too different from $a$, we re-ran the bisection algorithm using more sample points and a longer burn-in time. 

\subsection{Rejection statistics and projection diagnostics} \label{sec:reject}

\begin{figure}
\includegraphics[width=\textwidth]{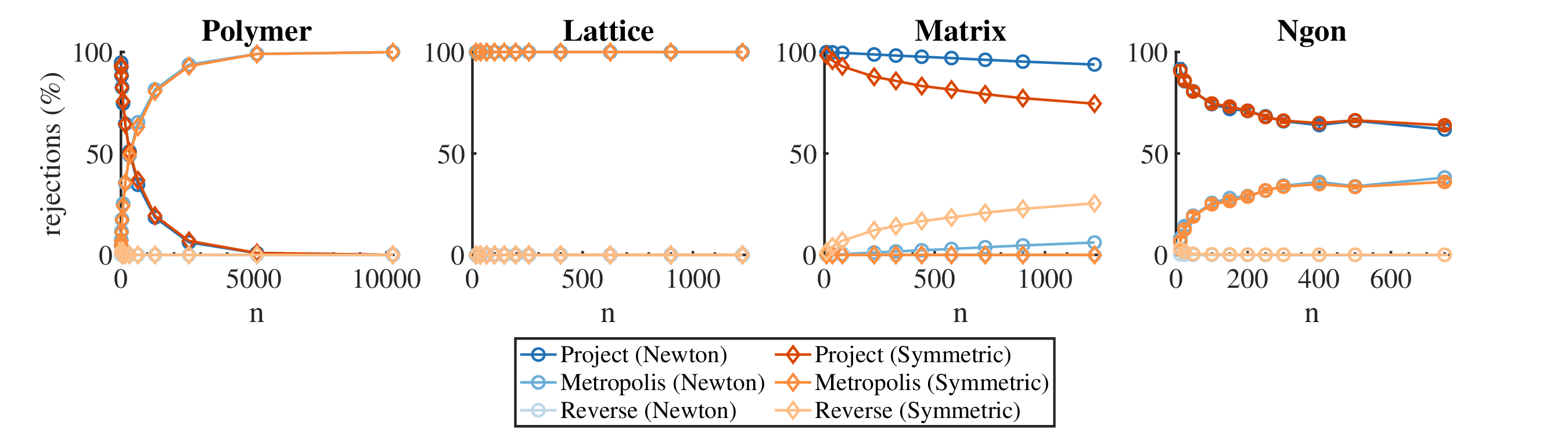}
\caption{
Rejection statistics for each example and each method (traditional Newton / symmetric Newton), as a function of $n$. Proposals may be rejected because the forward projection fails (Project), because the Metropolis-Hastings criterion is not satisfied (Metropolis), or because the reverse projection fails or produces a point different from the starting point (Reverse). 
The Polymer, Lattice, and Ngon have essentially 0 Reverse rejections; for these examples traditional and symmetric Newton have virtually identical statistics of each type. The Lattice has 0 Project rejections. 
Data for the Ngon is an average over 3 independent examples; the one-standard deviation error bars are smaller than the markers.
}\label{fig:rej}
\end{figure}

We next examine the rejection statistics and projection diagnostics, to see if there are notable differences between the methods and examples. 
Figure \ref{fig:rej} shows the fraction of each type of rejection, for each method: either the projection step failed to converge, or it was rejected in the Metropolis-Hastings step, or the reverse projection failed to converge. In most examples, at smaller $n$, rejections are mainly due to projections, while at larger $n$, Metropolis rejections become more important.  The Lattice was an exception, with only Metropolis rejections and no Projection rejections; this is because $\sigma$ was small to accommodate a strongly varying energy.

It is notable that the Polymer, Lattice, and Ngon, have identical rejection statistics for traditional Newton and symmetric Newton. 
The Matrix, on the other hand, has slightly different rejection statistics for traditional Newton and symmetric Newton -- traditional Newton has mostly projection rejections, with a small number of Metropolis rejections, while symmetric Newton has fewer rejections in the forward projection step, and some rejections in the reverse projection step (an amount that increases with $n$), with virtually no Metropolis rejections. Therefore, symmetric Newton in this example achieves the same overall acceptance rate $a=0.25$, by distributing its rejections between forward and reverse projections. 

In theory, the Matrix should have no rejections in the Metropolis step, as was discussed in Section \ref{ex:matrix}. The reason there are Metropolis rejections in practice for traditional Newton, is that we are not generating samples which satisfy the constraints exactly, so the tangent steps $v_x,v_y$ do not have exactly the same norm. We verified that decreasing the tolerance \texttt{tol} in Newton's method decreases the number of Metropolis rejections for traditional Newton. It is interesting that symmetric Newton does not have a significant number of Metropolis rejections -- we hypothesize this is because proposals where $|v_x|,|v_y|$ are more different are also more likely to not be found by the sloppier symmetric Newton projection. 

In all examples except the Matrix with symmetric Newton, the fraction of rejections in the reverse projection step was tiny -- less than 0.3\% and decreasing with $n$ for nearly all examples except the Ngon, for which the fraction was 2.7\%, 1.5\%, 0.5\% for $n=12,24,48$ respectively, and negligible thereafter.  Therefore, it appears that the reverse projection step could be skipped in these examples without significantly affecting the stationary distribution, particularly at large $n$. 


\begin{figure}
\includegraphics[width=\textwidth]{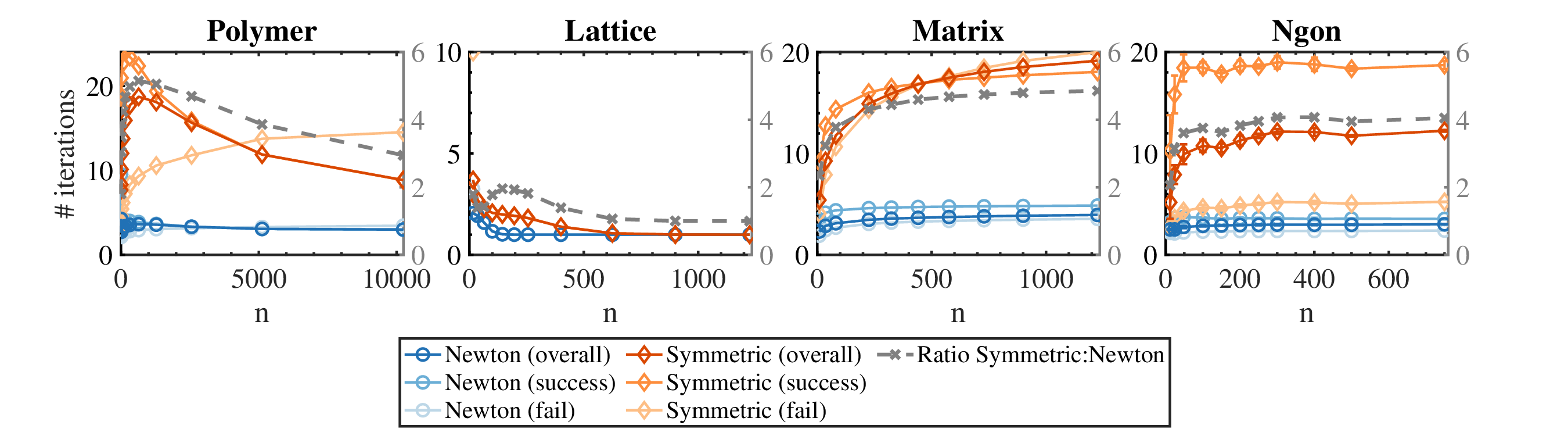}
\caption{Average number of iterations $n_{\rm iter}$ required for the projection method to converge, for traditional Newton and symmetric Newton, for different examples. The ratio symmetric:traditional is shown in grey on the right. Also shown are the average values of $n_{\rm iter}$ when the projection succeeds, and when it fails. 
Data for the Ngon is an average over 3 independent examples; the one-standard deviation error bars are generally smaller than the markers.
}\label{fig:niter}
\end{figure}

Since the rejection statistics are nearly identical for traditional and symmetric Newton for most examples, the algorithms are performing the same average amount of computation, except during the projection step. Let's study how the algorithms compare during this step. 

In each projection step, each algorithm performs a number $n_{\rm iter}$ of Newton or quasi-Newton iterations, during which there are various costly linear algebra operations.  
Figure \ref{fig:niter} shows the average $n_{\rm iter}$ for each method and example. Symmetric Newton generally requires more iterations than traditional Newton, but notably, not that many more: typically a factor of 3-5 times more, depending on the example. 
An exception is the Lattice at large $n$ where $n_{\rm iter} \approx 1$ for both methods, presumably because $\sigma$ is so small. 
Figure \ref{fig:niter} also shows the average values of $n_{\rm iter}$ conditional on a  projection succeeding and conditional on it failing. For traditional Newton these conditional values are similar, with a very slightly higher value for successful projections. For symmetric Newton, which conditional value is higher depends on $n$ and the example; though the difference is quite striking for the Ngon, where unsuccessful projections require far fewer iterations. This suggests our termination criterion for Newton's method is effective; we are not wasting time with projections that will not succeed.

\subsection{Matrix factorizations explain why symmetric Newton is faster than traditional Newton}\label{sec:matrixfac}

\begin{figure}
\includegraphics[width=\textwidth]{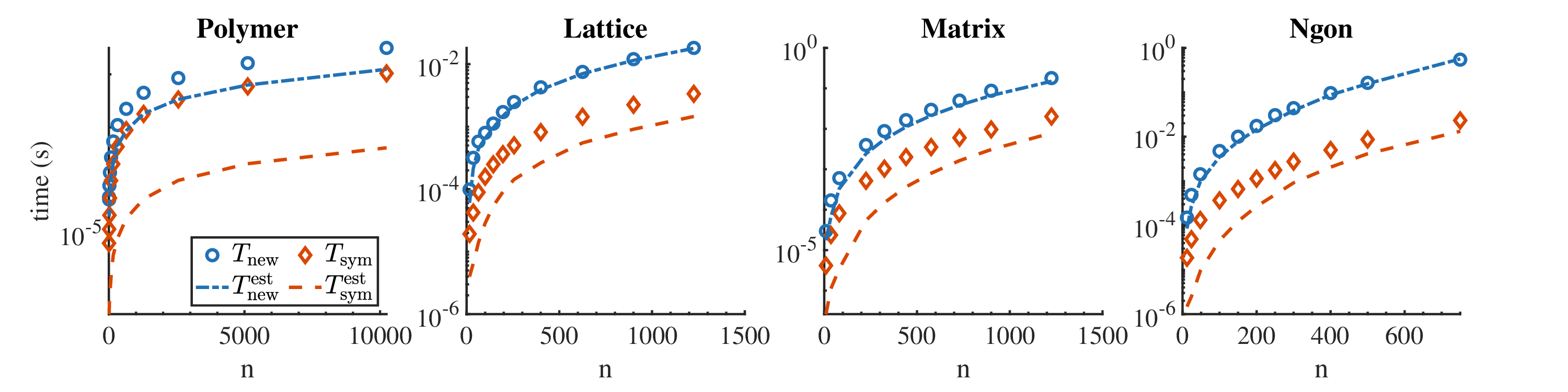}
\caption{Times to generate 1 point with each method, $T\new,T\sym$, and estimated times based on matrix factorizations, $T\new\est, T\sym\est$ from \eqref{Test}. 
}
\label{fig:timeest}
\end{figure}

\begin{table}
\centering
\begin{tabular}{l|ll|ll}
     & \multicolumn{2}{c|}{Newton} & \multicolumn{2}{c}{Symmetric} \\
     & small $n$ & large $n$ & small $n$ & large $n$ \\\hline
Polymer & 56\% ($n{=5}$) & 40\% (n{=}10,240)  & 4.6\% ($n{=5}$)  & 3.9\% ($n{=}10,240$)  \\
Lattice & 68\% ($n{=}36$) & 99\% ($n{=}1225$) & 15\% ($n{=}36$) &  43\% ($n{=}1225$)\\
Matrix & 40\% ($n{=}36$) & 83\% ($n{=}$1225) & 4.8\% ($n{=}$36) & 39\% ($n{=}1225$) \\
Ngon & 56\% ($n{=}24$) & 104\% ($n{=}740$) & 5.1\% ($n{=}24$) & 57\% ($n{=}750$) \\\hline
\end{tabular}
    \caption{Percentage of total computational time explained by matrix factorizations, $T\new\est/T\new$, $T\sym\est/T\sym$, for small $n$ and large $n$, for each example. We chose the smallest value of $n$ for the Lattice, Matrix, Ngon such that the ratios increased  monotonically as $n$ increased beyond that value. (The smallest values of $n$ generally had higher ratios.)}\label{tbl:timingest}
\end{table}

What sets the overall computational time of the algorithm? Let's count the number of linear algebra operations involving matrices that contribute to the computational time (those involving only vectors are expected to be insignificant). 
\begin{itemize}[nosep]
    \item Proposing a tangent step $v_x$: 2 matrix-vector multiplications, one forward-backward substitution.
    \item Projecting to obtain $y$: $n\iter$ times all of the following: one matrix-vector multiplication, one forward-backward substitution, and for traditional Newton, one LU decomposition and one Jacobian evaluation.
\end{itemize}
If $y$ is succcessfully proposed: 
\begin{itemize}[nosep]
\item Computing $v_y$: one matrix-matrix multiplication, one Cholesky decomposition, 2 matrix-vector multiplications, one forward-backward substitution.
\item If $y$ is not rejected by the Metropolis step: one reverse projection, involving the same steps as the forward projection. 
\end{itemize}
\medskip 


We expect the matrix factorizations to be the most expensive step, so we estimate the amount they contribute to the total computational time. Let $T\new, T\sym$ be the average computational time to generate 1 new point for each of traditional Newton and symmetric Newton respectively, determined empirically in Section \ref{sec:timing}, and let $T\new\est, T\sym\est$ be estimates obtained by counting the time associated with matrix factorizations. These estimates are constructed from the times $T\lu, T\chol$ required to perform one LU factorization and one Cholesky factorization respectively, multiplied by the average number of times each such factorization occurs. In each step of the Markov chain, symmetric Newton requires an average of approximately $1-r\sym(1-a)$ Cholesky factorizations, where $r\sym$ ($r\new$ for traditional Newton) is the fraction of rejections in the forward projection step, so that $r\sym(1-a)$ is the total fraction of proposals rejected in the projection step, hence which don't proceed to a Cholesky decomposition of a proposal $y$.
Traditional Newton requires approximately $1-r\new(1-a)$ Cholesky factorizations, and approximately $(1+a)n_{\rm iter}$ LU factorizations, where $an\iter = 0.25n\iter$ is added to account for the reverse projection step, which occurs approximately once per point accepted. Therefore, we may estimate the computational times as 
\begin{equation}\label{Test}
    T\sym\est = (1-0.75r\sym)T\chol, \qquad 
    T\new\est = 1.25n\iter T\lu + (1-0.75r\new)T\chol. 
\end{equation}
We constructed these estimates by first estimating $T\chol, T\lu$, which we did for each example by constructing a matrix $A$ with the same sparsity pattern as $Q_x^TQ_x$, and repeatedly performing either a Cholesky decomposition or a LU decomposition of this matrix. We measured the average time to perform $10^4$ factorizations of each (and repeated for several random matrices to ensure the results were independent of the particular entries, which they were). 

Figure \ref{fig:timeest} shows the actual and estimated timings, $T\sym, T\new$ and $T\sym\est,T\new\est$ respectively, and Table \ref{tbl:timingest} shows the percentages explained by matrix factorizations, $T_{(\cdot)}\est/T_{(\cdot)}$, for small and large $n$. 
For all examples except the Polymer, the estimates \eqref{Test} are significantly smaller than the measured values for small $n$, but their agreement increases with $n$, until at large $n$ they explain most of the timing -- 80-100\% for traditional Newton, and 40-60\% for symmetric Newton. For the Polymer, the percentage explained by the estimates actually decreases with $n$, though slowly, and is approximately 40-50\% for traditional Newton, and 4\% for symmetric Newton. 

We believe the Polymer behaves differently because it has so few constraints, with such a simple sparsity structure ($Q_x$ has 9 nonzero elements per column, and in $Q_x^TQ_x$ the nonzero elements are almost all concentrated near the diagonal), that performing matrix factorizations is roughly comparable to the other linear algebra operations. There are several reasons why $T\sym\est > T\sym$ for the Ngon with traditional Newton at large $n$ -- perhaps the 3 realizations behave differently and our estimate ignores such correlations in statistics; perhaps the slight difference in $n\iter$ for successful and failed projections, which we haven't accounted for in \eqref{Test}, is playing a role.

For the Lattice, Matrix, and Ngon, our estimates \eqref{Test} show that at large $n$, the primary contribution to the total time of the algorithm are the matrix factorizations. This is why symmetric Newton is so much faster than traditional Newton -- because it significantly reduces the number of matrix factorizations. This agreement also allows one to generalize our findings to other implementations of the algorithm, e.g. different machine architectures, sparsity orderings, linear algebra routines, etc -- we expect matrix factorizations to still dominate given other choices, so one can estimate the total time of the algorithm using the time it takes to perform a matrix factorization, as in \eqref{Test}. (This model also uses $r\sym, r\new$, which can be dropped without changing the estimates significantly.) Furthermore, the model shows that improving the factorizations, e.g. by performing them in parallel, or by avoiding them altogether (an issue we discuss in Section \ref{sec:conclusion}), would have the greatest effect on timing. 

It is notable however that for symmetric Newton, matrix factorizations represent about 50\% of the timing at large $n$ (though this percentage would be higher if we explored even larger $n$), and much less for smaller $n$, meaning that we have made matrix factorizations equivalent to the other linear algebra operations for the values of $n$ explored. This suggests that even if we were able to avoid a matrix factorization, e.g. by using iterative methods to solve systems of equations instead of the direct methods we proposed in this paper, the effect on timing would not be significant, except for larger values of $n$ than we have explored here.


\subsection{Effective diffusivity as a function of $a$} \label{sec:diff}

\begin{figure}
\includegraphics[width=\textwidth]{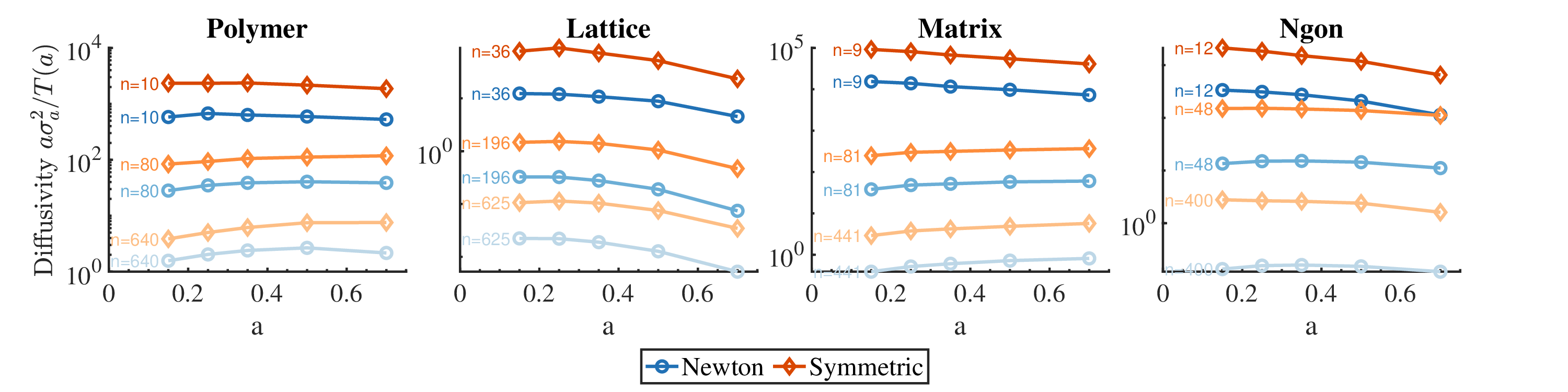}
\caption{Estimated diffusivity $a\sigma_a^2/\tau_c(a)$ as a function of average acceptance probability $a$, for several values of $n$. Traditional Newton is shown in shades of blue/circle markers; symmetric Newton is shown in shades of red/diamond markers. One sample was used for the Ngon. 
}\label{fig:diff}
\end{figure}

We repeated the experiments above with different values of $a$ to test the robustness of our observations, and found they remained qualitatively the same for all $a$. Quantitatively, the computational times changed, generally increasing weakly with $a$ by a factor of at most 2, except for the Lattice for which it was nearly constant. However, the ratio of times between traditional Newton and symmetric Newton remained roughly the same. The rejection statistics were also roughly similar, though with a smaller proportion of Projection rejections and a greater proportion of Metropolis rejections at larger values of $a$. This shift in rejection statistics could underlie the slight increase in computational time, since a decreasing proportional of forward projection rejections means an increase in the average number of matrix factorizations, since there are a greater fraction of successfully proposed $y$s. 

Which value of $a$ makes the method most efficient? We tested this by computing the effective diffusivity $D_{\rm eff}(a)$, assuming steps are small enough that diffusivity is estimated as for a random walk in a flat space by the averaged step size squared, divided by the mean computational time per step: 
\begin{equation}\label{diff}
    D_{\rm eff}(a) = a\sigma_a^2/T(a)
\end{equation}
Figure \ref{fig:diff} shows that the value of $a$ that maximizes $D_{\rm eff}(a)$ generally depends on $n$, as well as on the presence of an ambient energy function. For the three examples with no ambient energy function -- the Polymer, Matrix, and Ngon -- $D_{\rm eff}(a)$ is maximized at the smallest values of $a$ when $n$ is small, gradually transitioning to a maximum at large values of $a$ for larger $n$. For the Lattice, which has an ambient energy function, $D_{\rm eff}(a)$ is generally maximized near $a=0.25$; this is consistent with the theoretical predictions of $a=0.23$ (a value we didn't test) \cite{Gelman1996}. 

We may speculate why large $a$ is more efficient for large $n$, for the Polymer, Matrix, and Ngon. For these examples, at large $n$, $\sigma_a$ decreases only weakly with $a$. 
A small decrease in $\sigma$ therefore gives a large increase in $a$; this is mostly because the number of Metropolis rejections decreases -- apparently the difference in $|v_x|, |v_y|$ is felt more strongly at large $n$. 
For these values of $n$ it is advantageous to use a large acceptance ratio so the effective step size $a\sigma_a^2$ is larger; the slight increase in computational time and slight decrease in $\sigma$ do not appear to outweigh the gain in effective diffusivity \eqref{diff} due to increasing $a$. 
This issue does not apply to the Lattice, where $\sigma_a$ decreases more rapidly with $a$ at all values of $n$. In this example it is the energy function which determines the step size, and not the limits set by asking for approximate reversibility in the geometry.

It is also notable that $D_{\rm eff}(a)$ does not vary strongly with $a$ -- the most it varied over the range $a\in [0.15,0.7]$ for the Polymer, Matrix, Ngon was by a factor of 2 (a factor of 3 for the Ngon with small $n$), and by a factor of about 4 for the Lattice. 
Therefore, the choice of $a$ does not seem particularly important within reasonable limits.

\section{Discussion}\label{sec:conclusion}

We have introduced a numerical algorithm to implement a Markov Chain Monte Carlo method to sample probability distributions defined on manifolds, where the manifolds are defined by the level sets of constraint functions. The algorithm builds on the method proposed in \cite{Zappa:2018jy} by proposing a numerically efficient implementation of this method, which is easy to implement, and which scales well to problems with thousands of dimensions. The algorithm requires one matrix factorization, a Cholesky factorization of a large sparse matrix, per step, and it uses this factorization both to directly solve linear systems of equations, and to compute the determinant of a Jacobian matrix ($Q_x$) which is required in physical systems where the constraints are approximations for stiff forces. A key step in making the method maximally efficient is to use a quasi-Newton method to solve the nonlinear system of equations defining the manifold, by replacing the exact Jacobian of the equations with a symmetric approximation that doesn't change with iteration, an idea that was first explored in the molecular dynamics community \cite{Barth:1995vi}. 

We applied the algorithm to several examples with varying constraint structures, most of which were inspired by physics and materials science. Three out of four examples had no ambient energy function, which allowed us to focus purely on the geometrical aspects of the sampler.  
The computational complexity of the algorithm varies with the example, and is mainly determined by the complexity of performing a sparse matrix factorization. This in turn depends on the complexity of the constraints, with simpler constraints (such as for a polymer with fixed endpoints) having the smallest complexity, and dense constraints with no structure (such as for a random graph) having the highest complexity. The algorithm becomes infeasible when performing the requisite Cholesky factorization becomes infeasible. 

We estimated the fraction of time taken by the Cholesky factorization and found it was negligible for small system sizes, though increasing to about 50\% for the largest system sizes we considered (provided they had sufficiently complex constraints); presumably this fraction would continue to increase for even larger systems.
Therefore, for large system sizes a parallel implementation of the requisite linear algebra operations should be beneficial, though we haven't explored parallelization here. 

We estimated the effective diffusivity of the sampler (measured in units of space squared divided by computational time) as a function of the average acceptance probability $a$, and found that for problems with no ambient energy, the sampler was most efficient at small $a$ in low dimensions and at large $a$ in high dimensions. However, as the effective diffusivity did not vary by more than a factor of 2-3 with $a$, we conclude that the particular choice of $a$ is unimportant. 

For very large systems one might wish to use a method that doesn't require a matrix factorization. In fact, \emph{the only step in our algorithm for which the factorization is required} is computing the ratio of pseudodeterminants $|Q_y|^{-1}/|Q_x|^{-1}$ -- we know of no iterative way to compute this ratio. Other steps which used the factorization could  in principle be done using iterative methods, such as conjugate gradient to solve equations \eqref{proj1},\eqref{reversev},\eqref{QxQx},  or GM-RES to solve \eqref{QyQx}. Such iterative methods require only matrix-vector multiplications, and hence may be applied to extremely high-dimensional problems, however direct methods are generally preferable to iterative methods when available. 

Therefore, it is ultimately computing the ratio $|Q_y|^{-1}/|Q_x|^{-1}$ which appears to be the limiting factor in determining the sizes of the systems one may sample using the method we have introduced, provided one is interested in systems which require this factor. (In molecular dynamics this factor appears to be omitted, e.g. \cite{Barth:1995vi}, however if a system is quite floppy this should lead to nonnegligible errors in statistics \cite{vanKampen:1981ia}.)  Interestingly, one may avoid computing determinants by using a flow-based method to solve SDEs on manifolds \cite{zhang2020}, though it seems impossible to metropolize such a method. This flow-based method operates by taking an unconstrained step in the ambient space $x\to x+\xi$, and then projecting to the manifold by solving the flow $\dd{\phi}{t} = -\grad F(\phi)$, $\phi(0) = x+\xi$, where $F(x) = |q(x)|^2$; this projection implicitly accounts for the determinant factor. While solving the ODEs for the flow is expected to be slow, 
it would be interesting to test whether this is still faster than computing a matrix factorization in high dimensions. 

We anticipate the collection of numerical techniques developed here can be extended to other sampling and simulations algorithms on manifolds. We anticipate in particular that they will be useful when extended to a Monte Carlo method that samples a collection of manifolds (a stratification), which is necessary for example to simulate sticky particles where bonds are not constant but can break and form \cite{HolmesCerfon:2020dk}. Most ideas used here should apply, though new ones will be needed to efficiently handle the Jacobian factors associated with adding and subtracting bonds, which are not directly obtained from $Q_x$, or more generally in jumping to manifolds of lower or higher dimension.

\section*{Acknowledgements}
MHC would like to thank George Stadler and Jonathan Goodman for many helpful conversations, and Benedict Leimkuhler for introducing us to \cite{Barth:1995vi}. 
KX and MHC acknowledge support from Research Training Group in Modeling and Simulation funded by the National Science Foundation via grant RTG/DMS–1646339.
MHC acknowledges support from NSF-DMS-2111163, NSERC-GR026736, and the Alfred P. Sloan Foundation.


\bibliographystyle{plainnat}
\bibliography{SamplingHighDimensions}

\end{document}